\documentstyle[12pt]{article}
\def\a{\alpha}
\def\b{\beta}
\def\c{\chi}
\def\d{\delta}
\def\e{\epsilon}

\def\f{\phi}
\def\g{\gamma}

\def\j{\psi}

\def\l{\lambda}
\def\m{\mu}
\def\n{\nu}

\def\q{\theta}
\def\r{\rho}
\def\s{\sigma}
\def\t{\tau}

\def\F{\Phi}
\def\G{\Gamma}
\def\J{\Psi}

\def\S{\Sigma}

\def\inbar{\vrule height1.5ex width.4pt depth0pt}
\def\rlx{\relax\leavevmode}
\def\I{\leavevmode\hbox{\small1\kern-3.8pt\normalsize1}}
\def\openone{\leavevmode\hbox{\small1\kern-3.3pt\normalsize1}}
\def\Ione{\rlx{\rm 1\kern-2.7pt l}}
\font\cmss=cmss10
\font\cmsss=cmss10 at 7pt
\def\ZZ{\rlx\leavevmode
             \ifmmode\mathchoice
                    {\hbox{\cmss Z\kern-.4em Z}}
                    {\hbox{\cmss Z\kern-.4em Z}}
                    {\lower.9pt\hbox{\cmsss Z\kern-.36em Z}}
                    {\lower1.2pt\hbox{\cmsss Z\kern-.36em Z}}
               \else{\cmss Z\kern-.4em Z}\fi}
\def\Ik{\rlx{\rm I\kern-.18em k}}  
\def\IC{\rlx\leavevmode
             \ifmmode\mathchoice
                    {\hbox{\kern.33em\inbar\kern-.3em{\rm C}}}
                    {\hbox{\kern.33em\inbar\kern-.3em{\rm C}}}
                    {\hbox{\kern.28em\sinbar\kern-.25em{\rm C}}}
                    {\hbox{\kern.25em\ssinbar\kern-.22em{\rm C}}}
             \else{\hbox{\kern.3em\inbar\kern-.3em{\rm C}}}\fi}
\def\IP{\rlx{\rm I\kern-.18em P}}
\def\IR{\rlx{\rm I\kern-.18em R}}
\def\IN{\rlx{\rm I\kern-.20em N}}

\def\llsymbol#1{\@llsymbol{\@nameuse{c@#1}}}
\def\@llsymbol#1{\ifcase#1\or {}\or {'}\or {''}\or {'''}\or
   {''''}\or {'''''}\or  \else\@ctrerr\fi\relax}

\newcounter{contador}

\newcommand{\ol}\overline
\newcommand{\ti}\tilde
\newcommand{\wt}\widetilde
\newcommand{\wh}\widehat
\newcommand{\bv}\breve
\newcommand{\dg}\dagger

\newcommand{\C}{^{\mbox{\scriptsize c}}}
\newcommand{\sC}{\mbox{\scriptsize c}}

\newcommand{\QED}{QED$_{\mbox{\scriptsize 2+2}}$}
\newcommand{\Dddd}{$D$$=$$1$$+$$3$}
\newcommand{\DDdd}{$D$$=$$2$$+$$2$}
\newcommand{\Ddd}{$D$$=$$1$$+$$2$}

\newcommand{\AW}{{\mbox{\scriptsize AW}}}
\newcommand{\inv}{{\mbox{\scriptsize inv}}}
\newcommand{\aand}{\;\;\;\mbox{and}\;\;\;}

\newcommand{\be}{\begin{equation}}
\newcommand{\ee}{\end{equation}}
\newcommand{\bl}{\begin{eqnarray}&}
\newcommand{\el}{&\end{eqnarray}}
\newcommand{\bq}{\begin{eqnarray}}
\newcommand{\eq}{\end{eqnarray}}

\newcommand{\ts}{\textstyle}

\newcommand{\sx}{\sigma_x}
\newcommand{\sy}{\sigma_y}
\newcommand{\sz}{\sigma_z}
\newcommand{\sm}{{\s}^{\m}}

\newcommand{\ad}{{\dot\alpha}}
\newcommand{\bd}{{\dot\beta}}

\newcommand{\gm}{{\gamma}^m}

\newcommand{\uptad}{\widetilde\theta^{\dot\alpha}}

\newcommand{\qt}{\tilde\theta}
\newcommand{\qwt}{\widetilde\theta}
\newcommand{\ov}{\overline}
\newcommand{\pa}{\partial}
\newcommand\olj{\stackrel{~}{\overline\j}}
\newcommand\oljc{\stackrel{~}{\ol{\j}\C}}
\newcommand\swtpa{\stackrel{~}{\sl\wt\pa}}

\def\sl#1{\rlap{\hbox{$\mskip 1 mu /$}}#1}	
\def\ssl#1{\rlap{\hbox{$ {\scriptstyle /}$}}#1}
\begin{document}

\title{$N$$=$$1$ Super-$\tau_{3}$QED from Atiyah-Ward Space-Time}

\author{M. A. De Andrade\thanks{Internet e-mail: marco@cbpfsu1.cat.cbpf.br}
\hspace{0,1cm} and \hspace{0,1cm} O. M. Del Cima{\thanks{Internet e-mail:
oswaldo@cbpfsu1.cat.cbpf.br}} \\
Centro Brasileiro de Pesquisas F\'\i sicas (CBPF) \\
Departamento de Teoria de Campos e Part\'\i culas (DCP)\\
Rua Dr. Xavier Sigaud, 150 - Urca \\
22290-180 - Rio de Janeiro - RJ - Brasil.}

\date{}

\maketitle

\begin{abstract}
In this letter, we present the action for the massive super-{\QED}. A pair of
chiral and a pair of anti-chiral superfields with opposite $U(1)$-charges are
required. We also carry out a dimensional reduction {\it{\`a la}} Scherk from
($2$$+$$2$) to ($1$$+$$2$) dimensions, and we show that, after suitable
truncations are performed, the supersymmetric extension of the
${\tau}_{3}$QED$_{1+2}$ naturally comes out.
\end{abstract}

\vspace{0,3cm}

The idea of considering space-times with several time directions and indefinite
signature has deserved a great deal of attention since a self-dual Yang-Mills
theory in ($2$$+$$2$) dimensions {\cite{selfdym}} has been related to the
Atiyah-Ward conjecture {\cite{award}}: this theory might be the source for
various integrable models in lower dimensions, after appropriate dimensional
reductions are carried out.

More recently, Gates, Ketov and Nishino {\cite{gatesketnish1}} have pointed out
the existence of Majorana-Weyl spinors in the Atiyah-Ward space-time, and
$N$$=$$1$ self-dual supersymmetric Yang-Mills theories and self-dual
supergravity models have been formulated for the first time in this particular
space. Afterwards, $N$$=$$2$ self-dual super-Yang-Mills and $N$$=$$2$ and
$N$$=$$4$ self-dual supergravities have been formulated and these results have
been useful for the conjecture that $N$$=$$2$ superstrings have no possible
counter-terms at quantum level to all orders in string loops
{\cite{gatesketnish2}}.

Since over the past years 3-dimensional field theories {\cite{djt}} have been
shown to play a central r\^ole in connection with the behaviour of
4-dimensional theories at finite temperature {\cite{12}}, as well as in the
description of a number of problems in Condensed Matter Physics
{\cite{13,domavro,qedtau3}}, it seems reasonable to concentrate efforts in
trying to understand some peculiar features of gauge-field dynamics in 3
dimensions. Also, the recent result on the Landau gauge finiteness of
Chern-Simons theories is a remarkable property that makes 3-dimensional gauge
theories so attractive {\cite{finiteness}}. Very recently, this line of
investigation has been well-motivated in view of the possibilities of providing
a gauge-theoretical foundation for the description of Condensed Matter
phenomena, such as high-$T_{c}$ superconductivity {\cite{domavro}}, where the
QED$_{3}$ and ${\tau}_{3}$QED$_{3}$ {\cite{domavro,qedtau3}} are some of the
theoretical approaches that been forwarded as an attempt to understand more
deeply about high-$T_{c}$ materials.

Our purpose in the present letter is to show the relationship between massive
Abelian  $N$$=$$1$ super-{\QED} {\cite{trab1}} and $N$$=$$1$
super-${\tau}_{3}$QED, after a dimensional reduction {\it{\`a la}} Scherk is
carried out and suitable supersymmetry-preserving truncations are made in order
to suppress non-physical propagating modes in three dimensions.

The supersymmetric extension of the massive QED in {\Dddd} requires two chiral
superfields carrying opposite $U(1)$-charges {\cite{wesszugauge}}. On the other
hand, to introduce mass in the matter sector in {\DDdd}, without breaking
gauge-symmetry, we have to introduce four scalar superfields: a pair of chiral
and a pair of anti-chiral superfields; the supermultiplets of each pair exhibit
opposite $U(1)$-charges.

The massive Abelian $N$$=$$1$ super-{\QED} is described by the action :
\footnote{In this letter, we are adopting $\eta_{\m \n}$$=$$(+,-,-,+)$ for the
A-W space-time metric, $ds$$\equiv$$d^4xd^2\q$, $d\wt{s}$$\equiv$$d^4xd^2\qwt$
and $dv$$\equiv$$d^4xd^2{\q}d^2\qwt$, where $\q$ and $\qwt$ are Majorana-Weyl
spinors. Notice that we are using the following convention for charge
conjugation of all Weyl spinors: $\j\C$$\equiv$$i\sz\j^*$ and
$\wt\c\C$$\equiv$$i\sz\wt\c^*$. Our conventions for the supersymmetry covariant
derivatives are: $D_{\a}$$=$$\pa_{\a}$$-$$i\sl\pa_{\a \ad}\uptad$ and
${\wt{D}_{\ad}}$$=$$\wt{\pa}_{\ad}$$-$$i\sl\wt{\pa}_{\ad \a}\q^{\a}$. The
spinor indices are raised and lowered with the help of the following
antisymmetric tensors: $\e_{\a \b}$$=$$-\e^{\a \b}$$=$$i\sy$ and $\wt \e_{\ad
\bd}$$=$$-\wt \e^{\ad \bd}$$=$$-i\sy$. We use the abbreviations
$\q\s^\m\qwt$$\equiv$$\q^{\a}\s^{\m}_{\a \ad}\qwt^{\ad}$,
$\q\j$$\equiv$$\q^{\a}\j_{\a}$ and $\qwt\wt\c$$\equiv$$\qwt^\ad\wt\c_\ad$. For
more details about notation and conventions in {\DDdd}, see
ref.{\cite{trab1}}.}
\bq
S_{\inv}^{\AW}\!\!\!&=&\!\!\!-{1\over8}\left(\int{ds}\;W^{\sC} W +
\int{d\wt{s}}\;{\wt{W}}^{\sC} \wt{W}\right) +
\int{dv}\;\left(\J^{\dg}_+e^{4qV}{\wt X}_+ + \J^{\dg}_-e^{-4qV}{\wt X}
_-\right)+\nonumber\\&& +
\,i\,m\left(\int{ds}\; \J_+\J_--\int{d\wt{s}}\;{\wt X}_+ {\wt X}_-\right) +
\mbox{h.c.} \;\;\;\;\;, \label{massqed}
\eq
where $q$ is a dimensionless coupling constant and $m$ is a parameter with
dimension of mass. The $+$ and $-$ subscripts in the matter superfields refer
to their respective $U(1)$-charges. To build up the interaction terms, we have
used a mixing between the chiral and anti-chiral superfields (in order to
justify such a procedure, we refer to the works of Gates, Ketov and Nishino
{\cite{gatesketnish2}}). This mixed interaction term establishes that the
vector superfield be {\it complex}.

In the action of $N$$=$$1$ super-{\QED}, given by eq.(\ref{massqed}), the
chiral superfields $\J_+$ and $\J_-$ ($\wt D_\ad\J_{\pm}$$=$$0$), are defined
as follows :
\be
\J_{\pm}(x,\q,\wt{\q})=e^{i\qt\ssl{\tilde\pa}\q}\left[A_{\pm}(x)+
i\q\j_{\pm}(x)+i\q^2F_{\pm}(x) \right] \;\;\;\;, \label{psi+}
\ee
\be
\sl\wt\pa_{\ad \a}\equiv\wt\e_{\ad \bd}\wt\s^{\m \,\bd}_{~~~\a}\pa_\m \aand
\wt\s^{\m \,\bd}_{~~~\a}=\left(i\sx,-\sy,\sz,\I_2\right)^{\bd}_{~~\a}\;\;\;,
\ee
where $A_+$ and $A_-$ are complex scalar fields, $\j_+$ and $\j_-$ are Weyl
spinors, and $F_+$ and $F_-$ are complex auxiliary scalar fields. Moreover, the
anti-chiral superfields, ${\wt X}_+$ and ${\wt X}_-$ ($D_\a \wt
X_{\pm}$$=$$0$), are defined by :
\be
{\wt
X}_{\pm}(x,\q,\wt{\q})=e^{i\q\ssl{\pa}\qt}\left[B_{\pm}(x)+i\qwt\wt\c_{\pm}(x)+
i\qwt^{2}G_{\pm}(x) \right] \;\;\;\;, \label{chi+}
\ee
\be
\sl\pa_{\a \ad} \equiv\e_{\a \b}\s^{\m\,\b}_{~~~\ad}\pa_\m \aand
\s^{\m\,\b}_{~~~\ad}=\left(-i\sx,\sy,-\sz,\I_2\right)^{\b}_{~~\ad}\;\;\;,
\ee
where $B_+$ and $B_-$ are complex scalar fields, $\wt{\c}_+$ and $\wt{\c}_-$
are Weyl spinors, and $G_+$ and $G_-$ are complex auxiliary scalar fields.

In the Wess-Zumino gauge {\cite{wesszugauge}}, a complex {\it vector}
superfield, $V$, is written as
\be
V(x,\q,\qwt)=\frac12i\q\s^\m\qwt
B_\m(x)-\frac12\qwt^2\q\l(x)-\frac12\q^2\qwt\wt\r(x)-\frac14\q^2\qwt^2
D(x)\;\;\;\;, \label{supervector}
\ee
where $D$ is a complex auxiliary scalar field, $\l$ and $\wt\r$ are Weyl
spinors and $B_\m$ is a {\it complex} vector field.

The field-strength superfields, $W_\a$ and ${\wt W}_\ad$, defined by
\be
W_\a=\frac12{\wt D}^2D_\a V \aand
\wt{W}_\ad=\frac12{D}^2\wt{D}_\ad V \;\;,
\ee
respectively, satisfy the chiral and anti-chiral conditions,
$\wt{D}_{\bd}W_\a$$=$$0$ and $D_{\b}{\wt W}_\ad$$=$$0$ ; they read as below :
\bq
\left\{\begin{array}{l}
W_\a=e^{i\qt\ssl{\tilde\pa}\q}\left[
\l_\a+\q^\b\left(\e_{\a\b}{D}-\s^{\m\n}_{\a\b}G_{\m\n}\right)+
i\q^2\s^\m_{\a\ad}\pa_\m{{\wt\r}^{\,\ad}}\right] \\
\\
{\wt
W}_\ad=e^{i\q\ssl{\pa}\qt}\left[{\wt\r}_\ad+\qwt^\bd\left(\wt\e_{\ad\bd}{D}-
\wt\s^{\m\n}_{\ad\bd}G_{\m\n}\right)+
i\qwt^2\wt\s^\m_{\ad\a}\pa_\m{{\l}^{\,\a}}\right]
\end{array}\right.\;\;\;\;, \label{sstrenght}
\eq
\be
\s^{\m\n\,\a}_{~~~~\b}=\frac14(\s^{[\m}\wt\s^{\n]})^{\a}_{~\b} \aand
\wt\s^{\m\n\,\ad}_{~~~~\bd}=\frac14(\wt\s^{[\m}\s^{\n]})^{\ad}_{~\bd}\;\;\;,
\ee
where $G_{\m\n}$$=$$\pa_\m B_\n$$-$$\pa_\n B_\m$ is the field-strength.

By considering the superfields defined above, the following component-field
action stems from the superspace action of eq.(\ref{massqed}) :
\bq
S_{\inv}^{\AW}\!\!\!&=&\!\!\!\int{d^4x}\left\{
-\frac14i\right({\l\C}{\sl\pa}\wt\r+{\wt\r}\,\C{\wt{\sl\pa}}\l \left)-\frac18
G^{*}_{\m\n}G^{\m\n}-\frac14 D^*D \; +\right.
\nonumber\\
&&
-\;F_+^{*}G_+ - A_+^{*}\Box B_+ - {1\over2} i {\j _+\C} {\sl{\pa}} \wt{\c}_+ -
qB_{\m} \left({1\over2} i {\j_+ \C}   \sm \wt{\c}_+ + A_+^*{\pa}^{\m}B_+ -
B_+{\pa}^{\m}A_+^*  \right) + \nonumber \\
&&
+\;iq\biggl(A_+^*\wt{\c}_+ {\wt{\r}} +B_+\j_+\C {\l} \biggr) -  \left( qD+
q^2B_{\m} B^{\m}\right)A_+^*B_+ +\nonumber\\
&&
-\;F_-^{*}G_- - A_-^{*}\Box B_- - {1\over2} i {\j _-\C} {\sl{\pa}} \wt{\c}_- +
qB_{\m} \left({1\over2} i {\j_- \C}   \sm \wt{\c}_- + A_-^*{\pa}^{\m}B_- -
B_-{\pa}^{\m}A_-^*  \right) + \nonumber \\
&&
-\;iq\biggl(A_-^*\wt{\c}_- {\wt{\r}} +B_-\j_-\C {\l} \biggr) +  \left(
qD-q^2B_{\m} B^{\m}\right)A_-^*B_- +
\nonumber\\
&&\left.
+\;m \biggl(\frac12i\j_+\j_- - \frac12i\wt\c_+\wt\c_-
-A_+F_--A_-F_++B_+G_-+B_-G_+ \biggr)
\right\}+\mbox{h.c.} \;\;\;\;\;. \label{massaction}
\eq

Due to the fact that in massive super-{\QED} one must have two opposite
$U(1)$-charges to introduce mass at tree-level, and a complex vector superfield
in order to build up the gauge invariant interactions, we can read directly
from the action (\ref{massqed}), the following set of local
$U(1)_{\a}$$\times$$U(1)_{\g}$ trasformations :
\bq
&\left\{\begin{array}{l}
\d_{g} {A^*}_{\pm}={\pm}i q\b(x) {A^*}_{\pm}\\
\\
\d_{g} \j^{\C}_{\pm}={\pm}i q\b(x) \j^{\C}_{\pm} \\
\\
\d_{g} {F^*}_{\pm}={\pm}i q\b(x) {F^*}_{\pm}
\end{array}\right. \aand
\left\{\begin{array}{l}
\d_{g} B_{\pm}={\mp}i q\b(x) B_{\pm}\\
\\
\d_{g} \wt\c_{\pm}={\mp}i q\b(x) \wt\c_{\pm} \\
\\
\d_{g} G_{\pm}={\mp}i q\b(x) G_{\pm}
\end{array}\right.\;\;\;\;\;\;\;\;\;\;, \label{U(1)+-sym}
\eq
where $\b$$\equiv$$\a$$-$$i\g$ is an arbitrary infinitesimal complex function.
Notice that the complexified gauge transformations (\ref{U(1)+-sym}) read as
above because one has previously fixed to work in the Wess-Zumino gauge. As for
the gauge superfield components surviving the Wess-Zumino gauge, we have :
\bq
&\left\{\begin{array}{l}
\d_{g} \l=\d_{g} \wt\r=0\;\;\;,\\
\\
\d_{g} D=0 \;\;\;\;\mbox{and} \\
\\
\d_{g} B_{\m}= i\; \pa_{\m}\b \;\;\;.\label{gaugesupertrans}
\end{array}\right.
\eq

Therefore, in the Wess-Zumino gauge, the real part of $B_{\m}$ gauges the
$U(1)_{\g}$-symmetry with real gauge function $\g$, whereas its imaginary part
gauges the $U(1)_{\a}$-symmetry with real gauge function $\a$. The latter is an
ordinary phase symmetry, and we associate it with the electric charge. Indeed,
as we will see later on, the imaginary component of $B_{\m}$ will be taken as
the photon field. The parameter $\g$ generates a local Weyl-like invariance
{\cite{priv1}}. However, the vector field that gauges such a symmetry, namely
the real part of $B_{\m}$, will be suppressed in the process of dimensional
reduction, so that such an invariance will not leave track in \Ddd.

It should be emphasized that the mass bilinears in the action given by
eq.(\ref{massaction}) preserve the local
$U(1)_{\a}$$\times$$U(1)_{\g}$-symmetry, since their component matter fields
(fermion and scalar) carry opposite charges. Therefore, the opposite values of
the $U(1)$-charges play the central r\^ole of introducing mass to matter fields
without breakdown of the gauge-symmetry, similarly to what happens in {\Dddd}.

\begin{center}
\begin{tabular}{|c|c|} \hline
$ $\DDdd$ $ & $ $\Ddd$ $     \\ \hline\hline
$G_{\m\n}^*G^{\m\n}$ & $G_{mn}^*G^{mn}+2\pa_m\f^*\pa^m\f$\\\hline
$\j\C\sl\pa\wt\c$ & $\olj\g^m\pa_m\c$         \\ \hline
$\wt\c\C\swtpa\j$ & $\ol\c\g^m\pa_m\j$  \\ \hline
$iB_\m\j\C\s^\m\wt\c$ & $iB_m\olj\g^m\c-\f\olj\c$  \\ \hline
$B_\m B\pa^\m A^*$ & $B_m B\pa^m A^*$  \\ \hline
$iA^*\wt\c\wt\r$ & $ A^*\ol{\c}\C\r$  \\ \hline
$iB\j\C\l $ & $- B\olj\l$  \\ \hline
$B_\m B^\m A^*B$ & $B_m B^m A^*B+\f^2 A^*B$  \\ \hline
$i\j_+\j_-$ & $-\oljc_+\j_-$  \\ \hline
$i\wt\c_+\wt\c_-$ & $\ol{\c}\C_+\c_-$  \\ \hline
\end{tabular}
\end{center}

\centerline{Table 1: Dimensional reduction rules from {\DDdd} to {\Ddd}.}

\vspace{0,5cm}

It is well-known that outstanding supersymmetric models with extended
supersymmetry are closely related to simple supersymmetries in higher
dimensions {\cite{scherk,sohnius}}. As we are interested in simple
supersymmetric models in {\Ddd}, since those should be more relevant for
applications in Condensed Matter Physics {\cite{13,domavro,qedtau3}}, we
concentrate our efforts to investigate which sort of model comes out after a
suitable compactification from Atiyah-Ward space-time to 3 space-time
dimensions is accomplished. Therefore, it will be interesting to carry out a
dimensional reduction{\footnote{One uses the trivial dimensional reduction
where the time-derivative, $\pa_3$, of all component fields vanishes,
$\pa_3{\cal F}$$=$$0$. Also, it was assumed that $B_{\m}$ is reduced in the
following manner: $B^\m$$=$$(B^m , \f)$, where $\f$ is a complex scalar field
and the 3-dimensional metric becomes $\eta_{m n}$$=$$(+,-,-)$. Note that, $\l$,
$\r$, $\j$ and $\c$ are now Dirac spinors in {\Ddd}.}} ({\it{\`a la}} Scherk
{\cite{scherk}}) of the massive $N$$=$$1$ super-{\QED}. Bearing in mind that
this procedure should extend the supersymmetry {\cite{scherk,sohnius}} to
$N$$\,>$$1$, truncations will be needed in order to remain with a simple
supersymmetry and to suppress unphysical modes, {\em{i.e.}} spurious degrees of
freedom coming from {\DDdd} dimensions.

To perform the dimensional reduction of the massive $N$$=$$1$ super-{\QED}
action (\ref{massaction}) to {\Ddd}, use has been made of the rules presented
in the Table 1. As a result, it can be directly found the following
3-dimensional action : {\footnote{For notations and conventions adopted in this
letter for {\Ddd}, see ref.{\cite{trab1}}.}}
\bq
S_{\inv}^{D=3}\!\!\!&=&\!\!\!\int{d^3\hat{x}}\left\{
-\frac14i\right({\ov\l}{\gm {\pa}_m}\r+{\ov\r}{{\gm {\pa}_m}}\l \left)-\frac18
\biggl(G^{*}_{mn}G^{mn}+2{{\pa}_m \f}^*{\pa}^m \f \biggr)-\frac14
D^*D\;+\right.
\nonumber\\
&&
-\;F_+^{*}G_+ - A_+^{*}\Box B_+ - {1\over2} i {\ov\j _+} {\gm {\pa}_m} {\c}_+ -
 qB_{m} \left({1\over2} i {\ov\j_+ } \gm {\c}_+ + A_+^*{\pa}^{m}B_+ -
B_+{\pa}^{m}A_+^*  \right) + \nonumber \\
&&
+\;{1\over2}q \f {\ov\j_+ } {\c}_+ + q\biggl(A_+^* \ov{\c}_+\C {\r} -B_+\ov\j_+
{\l} \biggr) -  \left( qD+q^2B_{m} B^{m}+q^2\f^2\right)A_+^*B_+ +\nonumber \\
&&
-\;F_-^{*}G_- - A_-^{*}\Box B_- - {1\over2} i {\ov\j _-} {\gm {\pa}_m} {\c}_- +
 qB_{m} \left({1\over2} i {\ov\j_- } \gm {\c}_- + A_-^*{\pa}^{m}B_- -
B_-{\pa}^{m}A_-^*  \right) + \nonumber \\
&&
-\;{1\over2} q\f {\ov\j_- } {\c}_- - q\biggl(A_-^* \ov{\c}_-\C {\r} -B_-\ov\j_-
{\l} \biggr) +  \left( qD-q^2B_{m} B^{m}-q^2\f^2\right)A_-^*B_- +\nonumber\\
&&\left.
-\;m \biggl(\frac12 \ov\j_+\C\j_- + \frac12\ov\c_+\C \c_-
+A_+F_-+A_-F_+-B_+G_--B_-G_+ \biggr)
\right\}+\mbox{h.c.} \;\;\;\;\;, \label{action3}
\eq
where, after dimensional reduction, the coupling constant $q$ acquires
dimension of (mass)$^{1\over2}$.

Analysing the 3-dimensional action given by eq.(\ref{action3}), it can be
easily shown that the spectrum will unavoidably be spoiled by the presence of
ghost fields, since the free sector of the action is totally off-diagonal.
Therefore, truncations are needed in order to remove the spurious degrees of
freedom, as well as to give rise to a simple supersymmetric action in {\Ddd}.
First of all, to make the truncations possible, we need to diagonalize the
whole free sector, in order that the ghost fields be identified.

The diagonalization is achieved by looking for suitable linear combinations of
the fields which yield a diagonal free action (\ref{action3}). After extremely
tedious algebraic manipulations, we find the following transformations which
diagonalize the action $S_{\inv}^{D=3}$ :
\begin{enumerate}
\item{gauge sector :}
\be
\l=\frac1{\sqrt{2}} \left(\wh\r + \wh\l \right) \aand \r=\frac1{\sqrt{2}}
\left(\wh\r - \wh\l \right) \;\;\;;
\ee
\item{fermionic matter sector :}
\be
\j_+=\frac1{\sqrt2}\left(\wh\j_+-{\wh\j_-}^{\rm c}+\wh\c_++{\wh\c_-}^{\rm
c}\right) \aand \j_-=\frac1{\sqrt2}\left(\wh\j_-+{\wh\j_+}^{\rm
c}+\wh\c_--{\wh\c_+}^{\rm c}\right) \;\;\;;
\ee
\be
\c_+=\frac1{\sqrt2}\left(\wh\c_++{\wh\c_-}^{\rm c}-\wh\j_++{\wh\j_-}^{\rm
c}\right)  \aand \c_-=\frac1{\sqrt2}\left(\wh\c_--{\wh\c_+}^{\rm
c}-\wh\j_--{\wh\j_+}^{\rm c}\right) \;\;\;;
\ee
\item{bosonic matter sector :}
\be
A_+=\frac1{\sqrt2}\left(\wh A_+-\wh B_+\right) \aand
A_-=\frac1{\sqrt2}\left(\wh A_--\wh B_-\right) \;\;\;;
\ee
\be
B_+=\frac1{\sqrt2}\left(\wh A_++\wh B_+\right) \aand
B_-=\frac1{\sqrt2}\left(\wh A_--\wh B_-\right) \;\;\;;
\ee
\be
F_+=\frac1{\sqrt2}\left(\wh F_++\wh G_+\right) \aand
F_-=\frac1{\sqrt2}\left(\wh F_-+\wh G_-\right) \;\;\;;
\ee
\be
G_+=\frac1{\sqrt2}\left(\wh G_+-\wh F_+\right) \aand
G_-=\frac1{\sqrt2}\left(\wh G_--\wh F_-\right)\;\;\;.
\ee
\end{enumerate}
On the other hand, to simplify the Yukawa-interaction terms (gaugino-matter
couplings), we find that following field redefinitions for the bosonic matter
sector are convenient :
\be
\wh A_+=\frac1{\sqrt2}\left(\bv A_+-\bv A_-^*\right) \aand
\wh A_-=\frac1{\sqrt2}\left(\bv A_+^*+\bv A_-\right) \;\;\;;
\ee
\be
\wh F_+=\frac1{\sqrt2}\left(\bv F_+-\bv F_-^*\right) \aand
\wh F_-=\frac1{\sqrt2}\left(\bv F_+^*+\bv F_-\right)\;\;\;.
\ee

By replacing these field redefinitions into the action (\ref{action3}), one
ends up with a diagonalized action, where the fields, $\f$, $\wh\r$, $\wh\c_+$,
$\wh\c_-$, $\wh B_+$ and $\wh B_-$ appear like ghosts in the framework of an
$N$$=$$2$-supersymmetric model. Therefore, in order to suppress these
unphysical modes, truncations must be performed. Bearing in mind that we are
looking for an $N$$=$$1$ supersymmetric 3-dimensional model (in the Wess-Zumino
gauge), truncations have to be imposed on the ghost fields, $\f$, $\wh\r$,
$\wh\c_+$, $\wh\c_-$, $\wh B_+$ and $\wh B_-$. To keep $N$$=$$1$ supersymmetry
in the Wess-Zumino gauge, we must simultaneously truncate the component fields,
$\wh G_+$, $\wh G_-$, $D$, $a_m$ and $\t$ {\footnote{The $a_m$ field is the
real part of $B_m$, since we are assuming $B_m$$=$$a_m$$+$$iA_m$. Also, as
$\wh\l$ is a Dirac spinor, it can be written in terms of two Majorana spinors
in the following manner: $\wh\l$$=$$\t$$+$$i\bv\l$.}} . The truncation of $\t$
is dictated by the suppression of $a_m$. Now, the choice of truncating $a_m$,
instead of $A_m$, is based on the analysis of the couplings to the matter
sector: $A_m$ couples to both scalar and fermionic matter and we interpret it
as the photon field in 3 dimensions.

After performing these truncations, and omitting the $(\widehat{\;\;\;})$ and
$(\bv{\;\;\;})$ simbols, we find the following action in {\Ddd} :
\bq
S_{N=1}^{\t_3{\rm QED}}\!\!\!\!&=&\!\!\!\!\int{d^3\hat{x}}\left\{
{\frac 12}i{\ov\l}{\gm {\pa}_m}\l-\frac14 F_{mn}F^{mn}\;+\right.
\nonumber\\
&&
-\;A_+^{*}\Box A_+ - A_-^{*}\Box A_- + i {\ov\j _+} {\gm {\pa}_m} {\j}_+ + i
{\ov\j _-} {\gm {\pa}_m} {\j}_- + F_+^{*}F_+ + F_-^{*}F_- + \nonumber \\
&& -\;q A_{m}\biggl({\ov\j_+ } \gm {\j}_+ - {\ov\j_- } \gm {\j}_- +
iA_+^*{\pa}^{m}A_+ - iA_-^*{\pa}^{m}A_- - iA_+{\pa}^{m}A_+^* +
iA_-{\pa}^{m}A_-^* \biggr) + \nonumber \\
&&
 -\;iq \biggl(A_+ \ov{\j}_+ {\l} - A_-\ov\j_- {\l} - A_+^{*} \ov{\l} {\j}_+ +
A_-^{*} \ov{\l} {\j}_- \biggr) +  q^2 A_{m} A^{m}\left(A_+^*A_+ + A_-^*A_-
\right) +\nonumber \\
&&\left.
-\;m \biggl( \ov\j_+\j_+ - \ov\j_- \j_- + A_+^{*}F_+ - A_-^{*}F_- + A_+F_+^{*}
- A_-F_-^{*} \biggr)
\right\} \;\;\;\;\;\;\;\;\;\;, \label{action3diag}
\eq
where it can be easily concluded that this is a supersymmetric extension of a
parity-preserving action, namely, ${\tau}_{3}$QED {\cite{qedtau3}}. However, to
render our claim more explicit, we are going next to rewrite
(\ref{action3diag}) in terms of the superfields of $N$$=$$1$ supersymmetry in 3
dimensions.

In order to formulate the $N$$=$$1$ super-${\tau}_{3}$QED action
(\ref{action3diag}) in terms of superfields, we refer to the work by Salam and
Strathdee {\cite{Salam}}, where the superspace and superfields in {\Dddd} were
formulated for the first time. Extending their ideas to our case in {\Ddd}, the
elements of superspace are labeled by $(x^m,\q)$, where $x^m$ are the
space-time coordinates and the fermionic coordinates, $\q$, are Majorana
spinors, $\q\C$$=$$\q$. {\footnote{The adjoint and charge-conjugated spinors
are defined by $\ol\j$$=$$\j^\dg \g^0$ and $\j\C$$=$$-C\ol\j^T$, repectively,
where $C$$=$$\sy$. The $\g$-matrices we are using arised from the dimensional
reduction to {\Ddd} are: $\g^m$$=$$(\sx,i\sy,-i\sz)$. Note that for any
spinorial objects, $\j$ and $\c$, the product $\ol\j\c$ denotes $\ol\j_a
\c_a$.}}

Now, we are ready to introduce the formulation of $N$$=$$1$
super-${\tau}_{3}$QED in terms of superfields. To begin with, we define the
complex scalar superfields with opposite $U(1)$-charges, $\F_+$ and $\F_-$ , as
\be
\F_\pm=A_\pm+\ol\q\j_\pm-\frac12\ol\q\q F_\pm \label{scalar3a} \aand
\F_\pm^\dg=A_\pm^*+\ol\j_\pm\q-\frac12\ol\q\q F_\pm^* \;\;\;\;,
\label{scalar3b}
\ee
where $A_+$ and $A_-$ are complex scalar fields, $\j_+$ and $\j_-$ are Dirac
spinors and $F_+$ and $F_-$ are complex auxiliary scalar fields.

In the Wess-Zumino gauge, the gauge superconnection, $\G_a$, is written as
\be
\G_a=i(\g^m\q)_aA_m+\ol\q\q\l_a \label{gauge3a} \aand
\ol{\G}_a=-i(\ol\q\g^m)_aA_m+\ol\q\q\ol\l_a \;\;\;\;, \label{gauge3b}
\ee
where $A_m$ is the gauge-field and $\l_a$ is the gaugino (Majorana spinor).

Defining the field-strength superfield, $W_a$, according to :
\be
W_a=\frac12{\ol D}_b D_a\G_b \;\;\;\;,
\ee
with superderivatives given by
\be
D_a={\ol\pa}_a-i(\g^m\q)_a\pa_m \aand {\ol
D}_a=\pa_a-i(\ol\q\g^m)_a\pa_m~\;\;\;,
\ee
it can be found that
\bq
W_a\!\!\!&=&\!\!\!\l_a+\S^{mn}_{~~~ab}\q_bF_{mn}-
\frac{i}2\ol\q\q\;\g^m_{~~ab}\left(\pa_m\l_b\right)
\label{strength3a} \\
\noalign{\hbox{and}}  \nonumber \\
\ol
W_a\!\!\!&=&\!\!\!\ol\l_a-\ol\q_b\S^{mn}_{~~~ba}F_{mn}+
\frac{i}2\ol\q\q\left(\pa_m\ol\l_b
\right)\g^m_{~~ba} \;\;\;\;, \label{strength3b}
\eq
where $\S^{mn}$$=$${\ts\frac14}[\gamma^m,\gamma^n]$ are the generators of the
Lorentz group in {\Ddd}.

The gauge covariant derivatives we are defining for the matter superfields with
opposite $U(1)$-charges, $\F_+$ and $\F_-$, are given by
\be
\nabla_a\F_\pm=\left(D_a\mp iq\G_a\right)\F_\pm
\aand\ol\nabla_a\F^\dg_\pm=\left(\ol D_a\pm iq\ol\G_a\right)\F^\dg_\pm
\;\;\;\;, \label{deriv3}
\ee
where $q$ is a coupling constant with dimension of (mass)$^{1\over2}$.

By using the previous definitions of the superfields, (\ref{scalar3a}),
(\ref{gauge3a}), (\ref{strength3a}) and (\ref{strength3b}), and the gauge
covariant derivatives, (\ref{deriv3}), we found how to build up the $N$$=$$1$
super-${\tau}_{3}$QED action, given by eq.(\ref{action3diag}), in superspace;
it reads :
\be
S_{N=1}^{\t_3{\rm QED}}\!=\!-\int\! d\hat v\left\{{\frac12}\ol
WW+(\ol\nabla\F_+^\dg)(\nabla\F_+)+(\ol\nabla\F_-^\dg)(\nabla\F_-)-
m(\F_+^\dg\F_+-\F_-^\dg\F_-)\right\} \,,
\label{superqedtau3}
\ee
where the superspace measure we are adopted is $d\hat v$$\equiv$$d^3\hat x
d^2\q$ and the Berezin integral is taken as $\int\!d^2\q
$$=$$-\frac14\ol\pa\pa$. Therefore, we finally show, by using the superspace
formulation (\ref{superqedtau3}), that the action (\ref{action3diag}) we have
found after a dimensional reduction {\it{\`a la}} Scherk, and some suitable
truncations of the massive $N$$=$$1$ super-{\QED}, is certainly the simple
supersymmetric version of ${\tau}_{3}$QED.

Our final conclusion is that the massive Abelian $N$$=$$1$ super-{\QED}
proposed in ref.{\cite{trab1}} shows interesting features when an appropriate
dimensional reduction is performed. The dimensional reduction {\it{\`a la}}
Scherk we have applied to our problem becomes very attractive, since, after
doing some truncations to avoid unphysical modes, the $N$$=$$1$
super-${\tau}_{3}$QED is obtained as a final result. In fact, the Atiyah-Ward
space-time shows to be very fascinating as a starting point to formulate models
to be studied in lower dimensions.

\subsection*{Acknowledgements}

\small

The authors are deeply indebted to Dr. J.A. Helay\"el-Neto for suggesting the
problem and for many exhaustive discussions. They express their gratitude to
Dr. S.P. Sorella, Dr. O. Piguet, Dr. L.P. Colatto for patient and helpful
discussions. Thanks are also due to our colleagues at DCP, in special to Dr.
S.A. Dias, for encouragement and to the Heads of CBPF-DCP, Prof. J.S. Helman
and Prof. J.J. Giambiagi, for providing a very good atmosphere to work in our
department. CNPq-Brazil is acknowledged for invaluable financial help.

\end{document}